\documentclass[a4paper,twoside,11pt]{article}
\usepackage{amsfonts}
\usepackage{amssymb}
\usepackage[utf8]{inputenc}
\usepackage{mathrsfs}
\usepackage{amsmath}
\usepackage[plain]{algorithm2e}
\usepackage{mathptmx}
\usepackage{cancel}
\DeclareMathAlphabet{\mathcal}{OMS}{cmsy}{m}{n}
\usepackage{rotating}
\usepackage{graphicx}
\usepackage{subeqnarray}
\usepackage{xcolor}
  
\usepackage[final]{pdfpages}
\usepackage{graphicx}
\usepackage{subfig}
\usepackage[font=small,labelfont=bf]{caption}
\usepackage{geometry}
\usepackage{graphicx}
\usepackage[sort&compress,numbers]{natbib}
\usepackage{enumerate}
\usepackage{wasysym}
\usepackage[textsize=footnotesize]{todonotes}
\usepackage{url}

\usepackage{bm}

\definecolor{darkblue}{rgb}{0,0,0.6}
\definecolor{darkred}{rgb}{0.6,0,0}

\usepackage[colorlinks=true, urlcolor=black, citecolor=black, linkcolor=black, hyperfootnotes=false]{hyperref}

\usepackage{authblk}

\usepackage{natbib}
\usepackage{graphicx}

\newcommand{\IT}{\textrm{IT}}
\newcommand{\NT}{\textrm{NT}}
\newcommand{\MF}{\textrm{MF}}
\newcommand{\tgt}{\textrm{tgt}}
\newcommand{\MM}{\textrm{MM}}

\newcommand{\F}{\textrm{F}}

\newcommand{\eq}{\textrm{fast}}

\usepackage{authblk}

\author[1,2]{Michele Vodret\footnote{\url{mvodret@gmail.com}}}
\author[1,3]{Iacopo Mastromatteo}
\author[1,3]{Bence T\'oth}
\author[1,2,3]{Michael Benzaquen}
\affil[1]{Chair of Econophysics \& Complex Systems, Ecole polytechnique, 91128 Palaiseau Cedex, France}
\affil[2]{Ladhyx, UMR CNRS 7646, Ecole polytechnique, 91128 Palaiseau Cedex, France}
\affil[3]{Capital Fund Management, 23-25, Rue de l’Université 75007 Paris, France}
\date{\today \vspace{-1cm}}
\title{Microfounding GARCH Models and Beyond:  A Kyle-inspired Model with Adaptive Agents}
\date{\today}

\begin{document}
\maketitle
\begin{abstract}
We relax the strong rationality assumption for the agents in the paradigmatic Kyle model of price formation, thereby reconciling the framework of asymmetrically informed traders with the Adaptive Market Hypothesis, where agents use inductive rather than deductive reasoning. 
Building on these ideas, we propose a stylised model able to account parsimoniously for a rich phenomenology, ranging from excess volatility to volatility clustering.  While characterising the excess-volatility dynamics, we provide a microfoundation for GARCH models. Volatility clustering is shown to be related to the self-excited dynamics induced by traders' behaviour, and does not rely on clustered fundamental innovations. 
 Finally, we propose an extension to account for the fragile dynamics exhibited by real markets during flash crashes. 
\end{abstract}
\providecommand{\keywords}[1]{\textbf{Keywords:} #1}


\tableofcontents

\setlength{\parskip}{\medskipamount}

\section{Introduction}
 
There exists a plethora of statistical models~\cite{Bachelier,bscholes,bmandelbrot,arch,garch86,Bacry-Muzy,Propagator,rough_vol} to account for price dynamics and stylised facts in financial markets~\cite{facts,zumbach}. While most of these models can be very useful for quantitative predictions, their formulation is not based on interacting agents, that is, they are not microfounded. An interesting research agenda consists in understanding how statistical models can be rationalised in terms of sound microfoundations. In particular, one asks what are the minimal hypotheses regarding agents' behaviour needed to properly microfound a given statistical model; agent-based models   are in fact well-known for their versatility~\cite{articleIoriABM}.

During the early years of this research agenda~\cite{O'Harabook}, microfounded models relied on the Rational Expectation Hypothesis (REH), i.e. agents are endowed with perfect knowledge of the market model and have unlimited and cost-less computing power. While the REH allows for mathematical tractability and model interpretability, it has severe limitations regarding actual predictive power~\cite{Shiller, Leroy-Porter, Shiller_nobel, lo2008efficient}. Moreover, the deductive reasoning implied by the REH contrasts with the inductive reasoning on which humans often rely in complex situations~\cite{Kahneman_tversky, Arthur}. In face of uncertainty, in fact, humans rely on pattern recognition, hypotheses formation, deduction using currently held hypotheses and replacement of hypotheses if needed; this leads to feedback effects, and consequently booms and bursts. Eventually, the Efficient Market Hypothesis (EMH), implied by the REH, is replaced with the formulation of the Adaptive Market Hypothesis (AMH)~\cite{shiller-feedback,lo2004adaptive,lo2008efficient}. 
Although in the long-time limit of some games with adaptive agents the REH is recovered~\cite{Blume,Dindo2020wisdom,giachini2021rationality}, this is not always the case~\cite{Bottazzicrowd} and, moreover, there is growing evidence that many stylised facts in financial markets cannot be captured within a framework where the REH holds~\cite{behavioral_hommes,Leroy2013CanRA,naturebouchaud}. Nevertheless, classic microfounded models can serve as a useful starting point for building more refined pictures of how financial markets work.

In his pioneering work, Kyle~\cite{kyle} proposed a highly stylised microfounded model for price formation: asymmetrically informed traders use rational expectations while interacting in the presence of noise trading. The corresponding rational equilibrium brings price impact, namely the fact that trades induce price jumps~\cite{Propagator}. 
Notwithstanding, the Kyle model predicts that price volatility is smaller than that of the fundamental price and it is time independent:  if the fundamental price is interpreted as the efficient price, these findings are in strong contrast with empirical observations;  price volatility is actually much larger than that related to the fundamental price by a factor $\sim$ 5~\cite{Shiller,Leroy-Porter, Blackok}, and exhibits intricate statistics with clustering and power law tails~\cite{facts,micciche2002volatility}. Note that while the linear Kyle framework can be modified by considering a risk-averse market maker~\cite{Subrahmanyam},  an unrealistically high risk-aversion parameter~\cite{Leroy2013CanRA} is needed in order to match empirical estimates.

The intermittent dynamics of price volatility can be accounted for by means of statistical descriptions, such  as  GARCH models~\cite{garch86}. While GARCH models are compatible with volatility clustering and power law tails, they are not order-driven models~\cite{O'Harabook} and hence they cannot account for price impact. Moreover, no connection with the concept of fundamental price is given, and thus excess volatility is hardly definable. As a final note, since these statistical models are not microfounded, they leave the question of `why do large price fluctuations cluster in time?' without a formal answer.
The classic explanation for volatility clustering is stated by Engle, the doctoral advisor of Bollerslev, the author of GARCH models, in his Nobel prize lecture~\cite{Nobel_engle}: `So at a basic level, financial price volatility
is due to the arrival of new information. Volatility clustering is simply clustering of information arrivals. [...]'. This explanation is in line with EMH, which assumes that all the available information is encoded in the price. 
However, starting from the work of Cutler, Poterba and Summers in 1988~\cite{Cutler_poterba_summers}, there is growing evidence that fundamental innovations only account for a small fraction of price jumps~\cite{Bouchaud_news,jiang}, and clusters of jumps~\cite{Marcaccioli}. Thus, an alternative explanation is needed for volatility clustering.  The REH accounts neither for excess volatility, nor for the volatility clustering that is unrelated to fundamental innovations. 

Here we provide a microfounded explanation for excess volatility and volatility clustering without the need to assume highly risk-averse agents, or that fundamental innovations break in clusters. Instead, we consider agents in a Kyle setup who adapt their strategies over time~\cite{Arthur,Hommes}, in line with the AMH. We suppose that traders adapt their strategy because the environment is changing, namely, the noise trade level fluctuates over time. In this way, we recover a stationary regime in which the market maker sets the price according to temporarily fulfilled expectations or beliefs, which, in turn, give way to different beliefs when they cease to be fulfilled, and so on and so forth. As we shall see, these minimal ingredients introduce feedback, and thus account for extreme events such as flash crashes. In doing so, we provide a microfoundation for GARCH models as well. We mention that a framework similar to ours  has been recently proposed in Ref.~\cite{yifan} and \cite{inoua}, and previously in Ref.~\cite{kirman_kesten}; however, to our knowledge, this is the first model that accounts for volatility clustering, excess volatility, price impact and, therefore, an interplay between trades and prices which can ignite liquidity crises and flash crashes.

    The paper is organised as follows. In Section~\ref{sec:the_model} we present the model. Section~\ref{sec:results} contains our main results, that is an analytical characterisation of the dynamics in a simplified yet realistic limit, together with a numerical investigation of the intermittent price dynamics. In Section \ref{sec:extensions} we present two interesting extensions of the model, accounting for a risk-averse liquidity provider, and a cost-averse liquidity taker. Finally, in Section~\ref{sec:conclusion} we discuss our findings.

\section{Model -- Evolving market conditions and adaptive agents}
\label{sec:the_model}

Consider three agents, two liquidity takers and one liquidity provider. They trade a single security over multiple trading rounds. At each trading round $t = 1,2,\dots$,  the liquidity takers first build their own demands: the informed trader (IT) knows the fundamental price $p^\F_t$ of the security, and exploits this private information to make profits, while the noise trader (NT) trades for exogenous reasons. The liquidity provider, therein called market maker, filters the information about the fundamental price from the excess demand created by the liquidity takers, reflecting it into the price $p_t$. 

The market conditions are specified by the statistics related to the signal and the noise process, respectively the fundamental price $p^\F_t$ and the noise trader order flow $q^\NT_t$. These are modelled as Gaussian processes with zero mean and variance respectively given by $\omega^2$  and $\epsilon^2_t$, where
\begin{equation}
    \label{eq:NT_volatility}
    \epsilon^2_t = \epsilon^2+\delta\epsilon^2_t.
\end{equation}
The fluctuations $\delta\epsilon^2_t$ follow an Auto-Regressive process of order one (AR(1)) with zero mean, volatility $\delta\epsilon^2\ll \epsilon^2$ and typical timescale  $\tau_{\NT}$; in formulas:
\begin{equation}
\label{eq:errrorAR1}
    \delta\epsilon^2_t = \exp(-1/\tau_\NT)\delta\epsilon^2_{t-1}+\eta_t\sqrt{1-\exp(-2/\tau_\NT)},
\end{equation}
where $\eta_t$ is a Gaussian process with zero mean and volatility given by $\delta\epsilon^2$. Note that although the fluctuations $\delta \epsilon^2_t$ can be negative, we are considering them small enough ($\delta \epsilon^2 \ll \epsilon^2$) such that the overall noise trade variance $\epsilon^2_t$ is always positive.
Other choices can be made for the  structure of noisy order flow volatility fluctuations $\delta\epsilon_t$ that will not change qualitatively our results on intermittent volatility dynamics. Moreover, keeping the fundamental price volatility $\omega$ fixed while considering a time varying noisy order flow volatility $\epsilon_t$, is a matter of choice for the modeler, and will not change our results; in fact, the crucial point to obtain an intermittent dynamics for price volatility, as we shall see, is that the market conditions evolve through time.  However, we are implicitly assuming that the volatility of the fundamental price varies slower than that of noisy order flow. This seems to be a sound choice: fundamental price volatility   varies slowly, consistent with the slow dynamics of the fundamental price, while noisy order flow fluctuations change rapidly, reflecting the fast, yet persistent, `sentiment' dynamics in real markets.

The excess demand  is the sum of the noise trader's and informed trader's demands:
\begin{equation}
\label{eq:total_order_flow}
q_t = q^\NT_t+q^\IT_t.
\end{equation}
It is cleared by the market maker, who sets the price $p_t$ of the security reflecting the unknown fundamental price $p^\F_t$. This is done by filtering out from the only observable the market maker has access to, specifically, the excess demand $q_t$, the information about the fundamental price injected by the informed trader. To do so, the market maker needs a prior about the liquidity takers' strategies. 
The outcome of the market maker's decisions is a price $p_t$ given, at each timestep $t$, by:
\begin{equation}
\label{eq:linear_price_impact}
    p_t = \Lambda_t q_t,
\end{equation}
where $\Lambda_t$ is the price impact function. In what follows, we explain how Eqs~\eqref{eq:total_order_flow} and \eqref{eq:linear_price_impact} are microfounded in terms of agents' strategies. 

The informed trader knows that the price is set at each step by Eq.~\eqref{eq:linear_price_impact}. Moreover, past prices and excess demands are public. This implies that at the beginning of each trading round $t$, the informed trader knows the price impact function realised at the previous step. Accordingly, the informed trader adapts his strategy over time by observing the evolving market conditions, which he captures via the evolving price impact function. We assume that the informed trader's best estimate for the one-step-ahead price impact function is the last observed one, which is a plausible heuristic rule.
The informed trader is modelled as a risk-neutral utility maximiser, implying that his strategy, at each trading round $t$, reads~\cite{kyle}: 
\begin{eqnarray}
\label{eq:IT_strategy}
q^\IT_t &=& \frac{p^\F_t}{2\Lambda_{t-1}}.
\end{eqnarray}

The market maker knows about the liquidity takers' strategies, but he can only observe the realized excess demand $q_t$. Therefore,  the market maker does not know the volatility of the fundamental price and of non-informed (or noise) trades: the market maker believes that these are  $\hat\omega_t$ and $\hat\epsilon$, respectively. We are therefore assuming that the market maker does not update his belief about noisy order flow volatility $\hat\epsilon$, while he updates his belief about fundamental price volatility.\footnote{The idea, which will be formalised in what follows, is that the market maker revises his own belief about fundamental price volatility such that the price volatility expectation matches the price volatility estimate constructed from past price history. We shall see that this implies a feedback loop between past and future price volatility leading to the volatility clustering effect observed in empirical data.} Note that the market maker's beliefs are denoted by hatted symbols, at variance with the ground-truth parameters that characterise the market conditions.
The market maker is modelled as a risk-neutral, expected utility maximiser. This implies that the price at step $t$ is set to be the optimal estimator of the fundamental price $p^\F_t$, given the beliefs $\{\hat\omega_t,\hat\epsilon\}$ and the functional form of the informed trader's strategy $q^\IT_t$. 
Accordingly, the price impact at each trading round $t$ reads:
 \begin{equation}
\label{eq:dynamical_syst_pre}
    \Lambda_t   = \frac{2 \Lambda_{t-1}\hat\omega_t^2}{\hat\omega^2_t+4\Lambda^2_{t-1}\hat\epsilon^2}.
\end{equation}
Note that the equation above has the same structure as the standard solution for the risk-neutral Nash equilibrium of the single step Kyle model~\cite{kyle,bouchaud2018trades}; however, the price impact function $\Lambda_t$ does not correspond to the real Nash equilibrium, due to the imperfect knowledge of the market maker about fundamental and noise trade volatilities.  

 We assume that $\hat\omega_t$ is a slowly varying function of $t$, meaning that the market maker's belief varies on timescales smaller than that at which trading occurs. As we shall see, the dynamics allows for timescale separation: fast dynamics, which take place on timescales over which the beliefs $\hat\omega_t$ do not change, and  slow dynamics, which take place on timescales over which the market maker revises his own beliefs. In the next subsections we describe these dynamics.

\subsection{Dynamics with constant belief}
\label{subsec:fast}

\begin{figure}
    \centering
    \includegraphics[scale = 0.25]{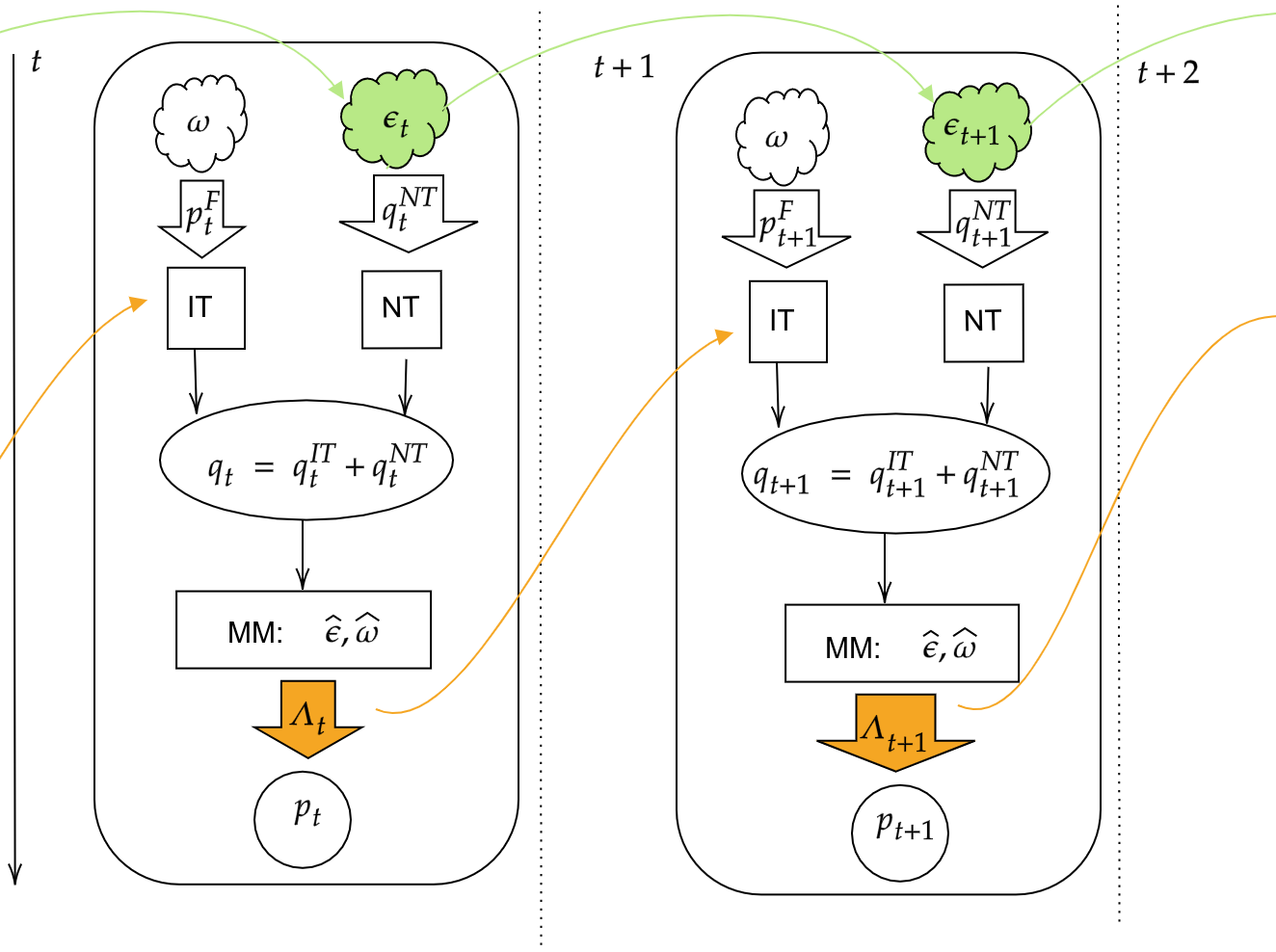}
    \caption{Dynamics with constant beliefs of the market maker, given by $\hat\epsilon$ and $\hat\omega$. Two trading rounds are explicitly shown, i.e., $t$ and $t+1$. The arrow of time in each trading round flows  from the top to the bottom. Consider the trading round $t$. First, a realization of the fundamental price $p^\F_t$ and of the noise trade $q^\NT_t$ are obtained from two indipendent Gaussian processes with zero mean and variance respectively given by $\omega$ and $\epsilon_t$. Then the liquidity takers create the excess demand $q_t$. Finally, the market maker clears the excess demand setting the price $p_t$ with a price impact function $\Lambda_t$, which becomes available information to the informed trader at the trading round $t+1$. Green and orange arrows refer to the dynamics related to the noise trade variance and to the price impact function, respctively given by Eqs.~\eqref{eq:NT_volatility},\eqref{eq:errrorAR1} and \eqref{eq:dynamical_syst_pre}.}
    \label{fig:diagram1}
\end{figure}

Consider a market maker's model that remains constant: $\hat\omega_t = \hat\omega$ for every $t$. In Fig.~\ref{fig:diagram1} we show schematically the dynamics with constant beliefs of the market maker.
The price impact dynamics, given by Eq.~\eqref{eq:dynamical_syst_pre}, admits a fixed point $\Lambda_\infty$ regardless of the initial condition $\Lambda_0$, and implicitly defines a relaxation timescale, given $\tau_{\eq}$, such that if $t\gg \tau_{\eq}$, then $\Lambda_t \sim \Lambda_\infty$. The choice for the subscript `fast' will become apparent shortly.
The price impact and the expected price variance at the fixed point read:
\begin{align}
\label{eq:pi_fixed}
    \Lambda_\infty &=  \frac{\hat\omega}{2\hat\epsilon},
    \\
\label{eq:vol_fixed}
    \hat\sigma^2_\infty &=    \frac{\hat\omega^2}{2},
\end{align}
which are reminiscent of the standard result of the Kyle model~\cite{kyle,bouchaud2018trades}, albeit these are calculated with market maker's beliefs.  
The relaxation timescale $\tau_{\eq}$ is obtained from a standard dynamical system argument and it is found to be equal to one trading round.\footnote{Let us mention here that $\tau_{\eq}$ can have a more interesting behaviour if the noise trader is cost-averse. We shall come back to this in Sec.~\ref{subsec:cost}} Therefore, the subscript `fast' relates to the fast equilibration dynamics of the price impact function. 

\subsection{Dynamics with belief revision}
\label{subsec:slow}

Consider a periodic updating procedure of the market maker's model, with period  $\tau_{\text{rev}}$. We refer to $k=0,1,2,\dots$ to denote the $k$'th update. In formulas, starting from $t=1$ and $k=0$, the fundamental price volatility   belief at time $t$ reads:
\begin{equation}
\label{eq:updating}
\hat\omega_t = \hat\omega_{k \tau_{\text{rev}}}, \quad \text{for} \quad k\tau_{\text{rev}}<t\leq (k+1)\tau_{\text{rev}}.
\end{equation}
According to the equation above, at the end of the trading rounds $t = k\tau_{\text{rev}}$ with $k\geq 1$, the market maker revises his belief about the fundamental price volatility.

\begin{figure}
    \centering
    \includegraphics[scale = 0.3]{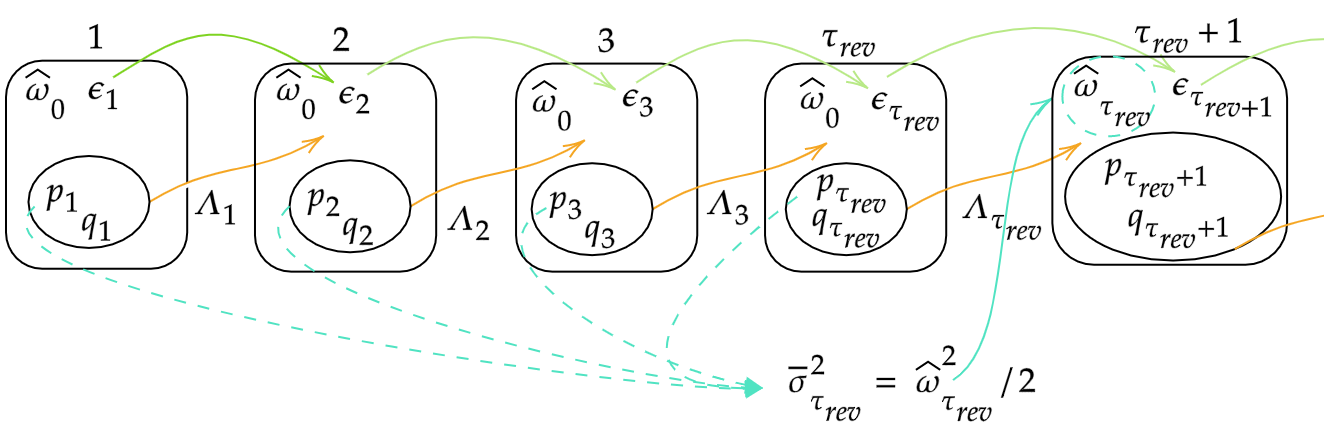}
    \caption{Dynamics with belief revision, starting from $t=1$. For the first $\tau_\text{rev}$ trading rounds the dynamics is represented, within each trading rounds, by boxes, as in Fig.~\ref{fig:diagram1}. The updating procedure at the end of step $\tau_\text{rev}$ is the only new feature: from the string of past $\tau_\text{rev}$ prices, the market maker computes the empirical estimate $\bar\sigma^2_{\tau_\text{rev}}$. From this estimate the market maker updates his belief about the fundamental price variance $\hat\omega^2_{\tau_{\text{rev}}}$ accordingly to Eq.~\eqref{eq:update_fundamental}. In addition to the green and orange arrows, already present in Fig.~\ref{fig:diagram1}, we show cyan arrows which represent the belief revision dynamics.}
    \label{fig:diagram2}
\end{figure}

The revision procedure comprises two steps. 
First, the market maker calculates an estimate of price volatility $\bar\sigma_{k \tau_{\text{rev}}}$ taking into account the last $\tau_{\text{rev}}$ recorded prices.
The updating timescale $\tau_{\text{rev}}$ controls the measurement error of the price volatility estimate: the larger  $\tau_{\text{rev}}$, the smaller the measurement error, that is the more precise is the estimate $\bar\sigma_{k \tau_{\text{rev}}}$.
Then, the market maker updates his fundamental price volatility belief $\hat\omega_{k\tau_\text{rev}}$ such that his current price volatility estimate $\bar\sigma_{k\tau_\text{rev}}$ equals the new long-time expected price volatility. Note that the relation between the expected long-time price volatility and fundamental price volatility  belief is given by Eq.~\eqref{eq:vol_fixed}. 
Accordingly, the updated belief about the  fundamental price variance satisfies the following condition:
\begin{equation}
\label{eq:update_fundamental}
    \frac{\hat\omega^2_{k\tau_{\text{rev}}}}{2} =  \bar\sigma^2_{k\tau_{\text{rev}}}.
\end{equation}
The dynamics with belief revision is shown schematically in Fig.~\ref{fig:diagram2}.

The heuristic rule given above coincides with $\hat\sigma_{k\tau_{\text{rev}}} = \bar\sigma_{k\tau_{\text{rev}}}$, if  $\tau_{\text{rev}} \gg \tau_{\eq}$  so that $\Lambda_t = \Lambda_\infty$ given by   Eq.~\eqref{eq:vol_fixed}. In what follows,  we shall refer to this regime as the sticky expectation one. In this regime the market maker is conservative, meaning that he slowly updates his belief; therefore, he constructs precise price volatility estimates, given that he uses a large number of past prices.

Now we have all the ingredients needed to understand why we choose to consider a time varying fundamental price volatility  belief $\hat\omega_{t}$: this choice allows to create a feedback loop between past prices and future price volatility. Specifically, past prices affect fundamental price volatility beliefs (see Eq.~\eqref{eq:update_fundamental}), which in turn affect prices via the price impact function (see Eqs.~\eqref{eq:linear_price_impact} and \eqref{eq:dynamical_syst_pre}).  This general feedback is well-known to be a feature of real markets~\cite{shiller-feedback}, where price volatility exhibits intermittent dynamics~\cite{facts}.

We anticipate here that the slow dynamics of updating beliefs can reach a stationary regime, which will be investigated in the next section. In this case, given an initial belief $\hat\omega_0$, there will be an associated relaxation timescale with which the stationary regime is achieved, denoted by $\tau_{\text{slow}} = k_{\text{slow}}\tau_\text{rev}$. Note that we used the subscript `slow' to distinguish the slow relaxation timescale, from the fast relaxation one denoted by `fast', related to the market maker's belief and the price impact function, respectively.

\section{Results}
\label{sec:results}

\subsection{Sticky expectation regime}

In this section we show that in the sticky expectation regime ($\tau_\text{rev}\gg \tau_\eq$) the model simplifies, allowing for an analytical characterisation of price volatility dynamics. Later, we elucidate the connection to GARCH models.

\subsubsection{Kesten dynamics}
\label{sec:kesten}

The timescale with which the price impact relaxes to its long-time limit $\tau_\text{fast}$ is of order one, as  stated in Sec.~\ref{subsec:fast}; therefore, we construct the sticky expectation regime as follows:
\begin{equation}
\label{eq:scaling_limit}
\frac{\tau_\text{fast}}{\tau_{\text{rev}}} \rightarrow 0,  \quad 
\frac{\tau_{\NT}}{\tau_{\text{rev}}} = r.
\end{equation}
The first condition implies that the updating timescale $\tau_{\text{rev}}$ is way larger than the equilibrium timescale $\tau_\text{fast}$ of the price impact dynamics. This allows to compute analytically the  price volatility estimate $\bar\sigma_{k \tau_{\text{rev}}}$, thanks to two simplifications. First, it allows to replace, in the calculation for price volatility estimate $\bar\sigma_{k \tau_\text{rev}}$, the price impact function with its fixed point value given by Eq.~\eqref{eq:pi_fixed}; note that one has to replace in Eq.~\eqref{eq:pi_fixed} $\hat\omega$ with its time dependent value according to Eq.~\eqref{eq:updating}. Second, since the price volatility estimate is constructed with a large number of observations ($\tau_{\text{rev}} \rightarrow \infty$), we can neglect the associated measurement error. Accordingly, we simplify the notation of price volatility expectation  $\bar\sigma_{k \tau_{\text{rev}}}$ by removing the bar symbol.  The second condition in Eq.~\eqref{eq:scaling_limit}  implies that noisy order flow volatility varies in the intervals $(k\tau_{\text{rev}}, (k+1)\tau_{\text{rev}}]$  if $r<\infty$. In the remaining part of the section, we measure time in units of $\tau_{\text{rev}}$, leaving only the index $k$, introduced in Eq.~\eqref{eq:updating} to denote the trading rounds at which market maker's beliefs are updated. 

The excess volatility $\sigma_k/\omega$ dynamics can be analytically characterised in the sticky expectation regime  starting from Eq.~\eqref{eq:update_fundamental} and using Eqs.~\eqref{eq:total_order_flow}, \eqref{eq:linear_price_impact}, \eqref{eq:IT_strategy} and \eqref{eq:pi_fixed}. With the simplified notation detailed above, the dynamics for the excess variance reads:
\begin{equation}
    \label{eq:empirical_vol}
    \cfrac{\sigma^2_{k}}{\omega^2} = \frac{1}{4}+  \cfrac{\epsilon^2_k}{2\hat\epsilon^2}\cfrac{\sigma^2_{k-1}}{\omega^2}.
\end{equation}
Therefore, the excess variance is a Kesten process~\cite{kesten}, i.e., a stochastic multiplicative process repelled from zero. The Kesten dynamics implies the possibility of having an intermittent dynamics for price volatility  together with large price volatility fluctuations captured by power law behavior.
 In fact, the dynamics depends crucially on the mean value of the stochastic multiplicative factor $ {\epsilon^2_k/(2\hat\epsilon^2)}$. In the case where $\delta\epsilon^2_k$ are iid, if $\langle {\epsilon^2_k/(2\hat\epsilon^2)} \rangle>1$, the Kesten process diverges and no stable distribution is reached. Converesely, if $\langle {\epsilon^2_k/(2\hat\epsilon^2)} \rangle<1$, the Kesten process is stable and approaches a limiting distribution for large times.\footnote{We will clarify what we mean by large times below, when we analyse the dynamics more precisely.}   This condition has an implication for the market maker's belief about noisy order flow volatility:  the stationary regime can be reached if and only if the market maker's underestimation of the noise trade variance  $\hat\epsilon^2$ is smaller than a   critical threshold  $\hat\epsilon^2_c$. More details about this and about the generalization to the case of correlated $\delta\epsilon^2_k$ are given below. 
 The equation above, 
after some manipulation involving Eq.~\eqref{eq:NT_volatility}, can be rewritten as:
\begin{align}
    \label{eq:similar_garch}
    {\sigma^2_{k}} &= \left\langle {\sigma^2} \right\rangle+
    \frac{\epsilon^2}{2\hat\epsilon^2} \left({\sigma^2_{k-1}}-\left\langle \sigma^2 \right\rangle\right)+
    \frac{\delta\epsilon^2_k}{2\hat\epsilon^2}  {\sigma^2_{k-1}},
\end{align}
Accordingly, price volatility dynamics is  the sum of three terms: a constant long-time contribution $\langle \sigma^2\rangle$, a deterministic mean-reverting contribution and a stochastic one, which represents the update based on the last empirical observation, which in turn reflects the evolving market conditions, i.e. the noise trades fluctuations $\delta\epsilon^2_k$.

\subsubsection{Comparison to GARCH models}
\label{subsec:garc_comparison}

In the quantitative finance literature, GARCH models~\cite{garch86} are very well-known~\cite{mantegna1999introduction,bouchaud_risk}. These are statistical models constructed explicitly in order to capture the intermittent dynamics of price volatility. In these models, price changes $\delta p_k$ are modeled as the product between the equal time volatility and Gaussian iid random variables $\xi_k$ with zero mean and unit volatility. The simplest model of this class is completely characterised by Eq.~\eqref{eq:similar_garch}, with the substitutions $\frac{\epsilon^2}{2\hat\epsilon^2} = \alpha$ and $
    \frac{\delta\epsilon^2_k}{2\hat\epsilon^2} = g (\delta p^2_{k-1}-1)$, following the notation in Ref.~\cite{bouchaud_risk}, 
where $\alpha<1$ and $g > 0$. This model is coined GARCH$(1,1)$, since only the previous time price volatility ($\sigma_{k-1}$) and price changes ($\delta p_{k-1}$) are taken into account.  
Although the sticky expectation regime of our model is characterized by a GARCH-like structure, the model we set up has a richer structure, highlighted below.
\begin{itemize}
    \item In GARCH models no connection between market price and fundamental price is provided, at variance with our Kyle inspired model. In our model, excess volatility is the  variable of interest, and not price volatility by itself (see Eq.~\eqref{eq:empirical_vol}).
\item  Although the price volatilty in the sticky expectation regime of our model is of the GARCH type,  our model is able to provide a microscopic interpretation for each of the three terms which appear in  Eq.~\eqref{eq:similar_garch}. The first one, i.e., the long-time price variance $\langle\sigma^2\rangle$ in our model is related to the distance between the market condition and market maker's beliefs, as we shall see in Sec.~\ref{subsec:mean_excess}. Similarly, the second, i.e., the coefficient of mean-reversion is related to the ratio $\epsilon^2/(2\hat\epsilon^2)$. Finally, the  variability of the interaction term, in the sticky expectation regime of our model, is not due to the measurement error of price volatility, as in the GARCH$(1,1)$ model, but rather to the time variability of noisy order flow volatility: in the GARCH$(1,1)$ model, the price volatility estimate ($\delta p_k$) is calculated only with the previous price change, implying a sensible measurement error; conversely, in the sticky expectation regime of our model, price volatility estimate is constructed from the past $\tau_{\text{rev}} \rightarrow \infty$ prices, obtaining an estimate without measurement error. 
    \item  
    Individual returns, in the sticky expectation regime of our model, are not explicitly modeled, because only their volatility and the volatility of the noise trades affect the excess-volatility dynamics (see Eq.~\eqref{eq:empirical_vol}). This is not the case for the GARCH$(1,1)$ model, where the kurtosis of returns can be calculated~\cite{mantegna1999introduction,bouchaud_risk}. In order to address directly this quantity one must resort to the full model described in Sec.~2.
    \item The stochastic multiplicative factor $\delta p^2_k$ of GARCH models is uncorrelated, whereas the one in our model ($\delta\epsilon^2_k$) it is an AR(1) process. Accordingly, in the GARCH$(1,1)$ model, the Auto-Correlation function (ACF) of price variance is  a single decaying exponential~\cite{mantegna1999introduction}, with correlation timescale  $\tau_{\text{ACF}} = 1/|\log(\alpha)|$. The sticky expectation regime of our model recovers this result if $r\rightarrow 0$; as we shall see below, in the generic case of finite $r$, the ACF of price volatility is more complicated than that predicted by GARCH model. 
    
    \item The model we built allows also to make predictions on a non-observable quantity, namely the ratio between informed and non informed orders. Further details on this matter are given in Sec.~\ref{subsec:constan_market_conditions}. 
\end{itemize}

\subsection{Excess volatility}
\label{subsec:mean_excess}

In the following we characterise the excess volatility. First we consider the case with fixed noise trade variance, then we consider the case where it fluctuates. For each investigation, first the sticky expectation regime is considered as it is easily interpretable and manageable; later, we highlight the differences with the simulation of the model presented in Sec.~\ref{sec:the_model}, which are affected by the measurement error on the price volatility estimate of the market maker. In Appendix~\ref{app:how_to_simulate} we present a pseudo-code for the simulation.

\subsubsection{Static market conditions}
\label{subsec:constan_market_conditions}

Consider the case where the noise trade volatility does not fluctuate over time $\epsilon_k = \epsilon$. In what follows we refer to this case as the Mean Field (MF) regime. In this regime, the excess-volatility dynamics in the sticky expectation regime of our model, given by  Eq.~\eqref{eq:empirical_vol}, becomes a deterministic dynamical equation. 
In the stable regime ($\epsilon^2/(2\hat\epsilon^2)<1$), the excess volatility converges to a finite fixed point. The relaxation is exponentially fast with timescale 
\begin{equation}
\label{MF_timescale}
    \tau_{\text{slow}} =  \tau_\text{rev}\left(1-\frac{\hat\epsilon^2_c|_\MF}{\hat\epsilon^2} \right)^{-1},
\end{equation}
where $\hat\epsilon^2_c|_\MF = \epsilon^2/2$. Therefore, the timescale with which the stationary regime is achieved  $\tau_{\text{slow}}$
 diverges as the market maker underestimation $\hat\epsilon^2$ approaches the critical value $\hat\epsilon^2_c|_\MF$.
The mean excess volatility can be easily computed from Eq.~\eqref{eq:empirical_vol}, obtaining:
\begin{equation}
\label{eq:MF_excess}
    \left\langle \cfrac{\sigma^2}{\omega^2}\right\rangle_{\MF} = \frac{1}{4}\left(1-\frac{\hat\epsilon^2_c|_\MF}{\hat\epsilon^2}\right)^{-1}.
\end{equation}
The equation above implies that the more the market maker underestimates the level of noise trading, the more the fixed point for the mean excess volatility grows. This can be easily explained as follows: if the market maker underestimates the mean value of noisy order flow volatility, he will overestimate the price impact function, leading to excessively volatile prices. In the case where the fluctuations $\delta\epsilon_k$ are iid, we recover the MF result, because $\langle \epsilon^2_k \sigma^2_{k-1}\rangle =  \epsilon^2 \langle \sigma^2 \rangle$.

Interestingly, our model allows to predict the ratio between informed and non informed trades. In fact, in the MF limit of the sticky expectation regime of our model, one finds:
\begin{equation}
\label{eq:ratio_informed}
    \left\langle \cfrac{\epsilon^2}{\epsilon^2_{\IT}}\right\rangle_\MF = \cfrac{\epsilon^2}{\hat\epsilon^2} \left(2-\cfrac{\epsilon^2}{\hat\epsilon^2} \right)^{-1},
\end{equation}
where $\epsilon^2_\IT$ is the informed trade variance.

\paragraph{Numerical simulations}

\begin{figure}[t]
\centering    \includegraphics[scale = 0.6]{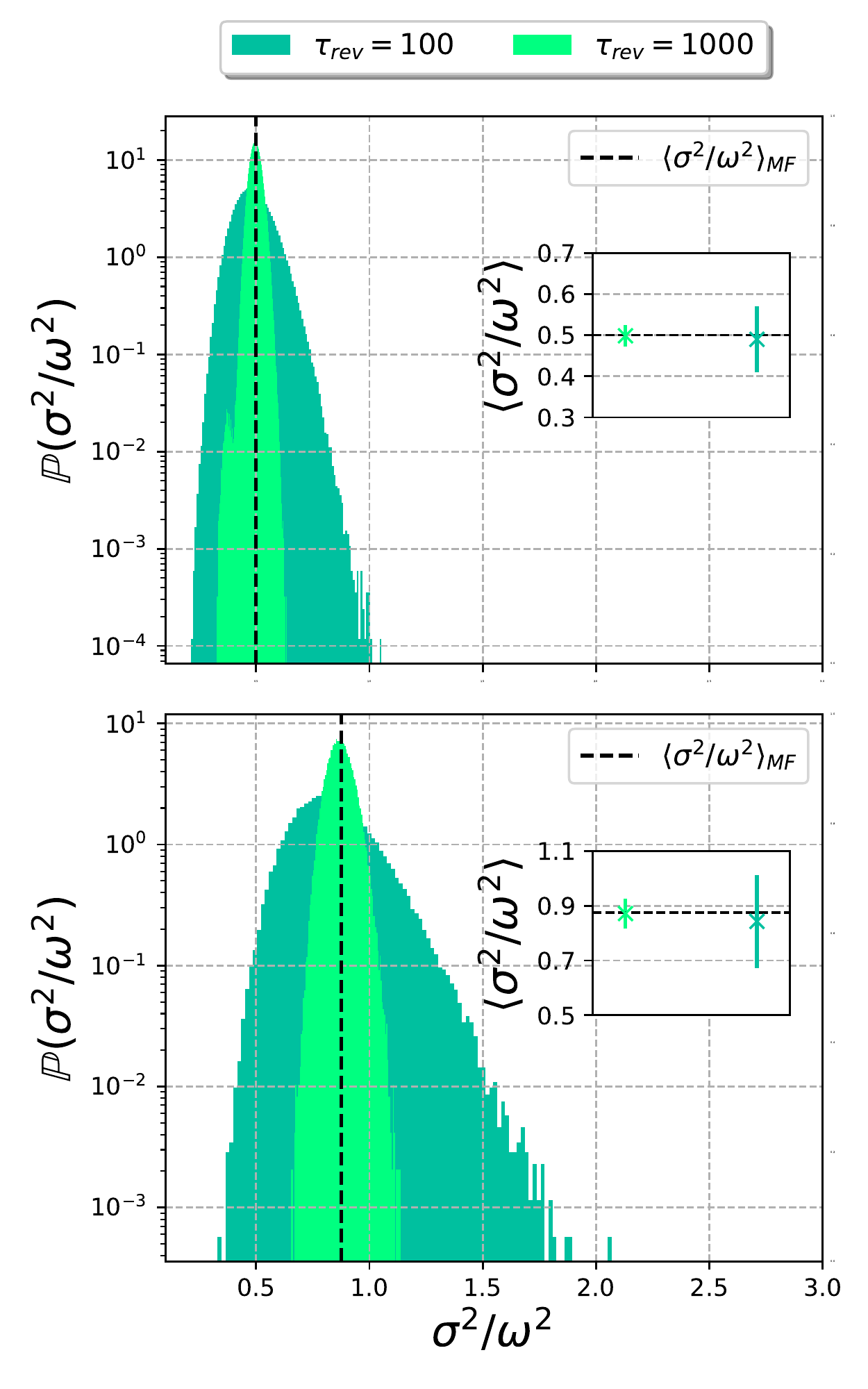}
    \caption{PDFs of excess variance with constant noise trade variance fluctuations ($\delta\epsilon^2 = 0$), varying the updating timescale $\tau_{\text{rev}}$. The larger the updating timescale, the more the excess-variance PDF is peaked around the mean value. (Top) Noise trade variance belief equal to true value. (Bottom) Noise trade variance belief lower than true value: $\hat\epsilon^2/\epsilon^2 = 0.7$. Black dashed vertical lines show the MF level of excess variance. Insets compare the MF level of excess variance with simulation outcomes (with error bars).}
    \label{fig:1}
\end{figure}

First, we analyze the case where the market maker knows perfectly the statistical properties of the noisy order flow volatility, i.e., $\hat\epsilon = \epsilon$; this is the case in which the REH holds. In the top panel of Fig.~\ref{fig:1} we show results about the  Probability Distribution Function (PDF) of excess variance for different values of the updating timescale  $\tau_{\text{rev}}$. There, one sees that  the larger $\tau_{\text{rev}}$ is, the more the excess variance is peaked around the strong rationality equilibrium value $\langle \sigma^2/\omega^2\rangle = 1/2$~\cite{kyle,bouchaud2018trades}. Fluctuations are induced by the measurement errors that affect the  price variance estimate. Note also that the larger $\tau_{\text{rev}}$, the more the distribution is peaked, since more past prices are taken into account to obtain the estimate for price variance. The inset shows in a more clear way that the mean excess variance is in agreement with the MF value of the sticky expectation regime that we analysed in the previous section.

Next, we consider the case where  the market maker underestimates the level of noisy order flow volatility, i.e., $ \hat\epsilon < \epsilon$. In this case, the excess variance is shifted to larger values, as we show in the bottom panel of Fig.~\ref{fig:1}. Note that fluctuations of the excess variance are more pronounced, since the underestimation of the noise trade level boosts the effect of the noise trade variance fluctuations.
The inset shows again that the mean excess variance is in agreement with the MF value of the sticky expectation regime (see Eq.~\eqref{eq:MF_excess}).

\subsubsection{Dynamic market conditions}
\label{subsec:dynamic_market}
\begin{figure}
    \centering
    \includegraphics[scale = 0.62]{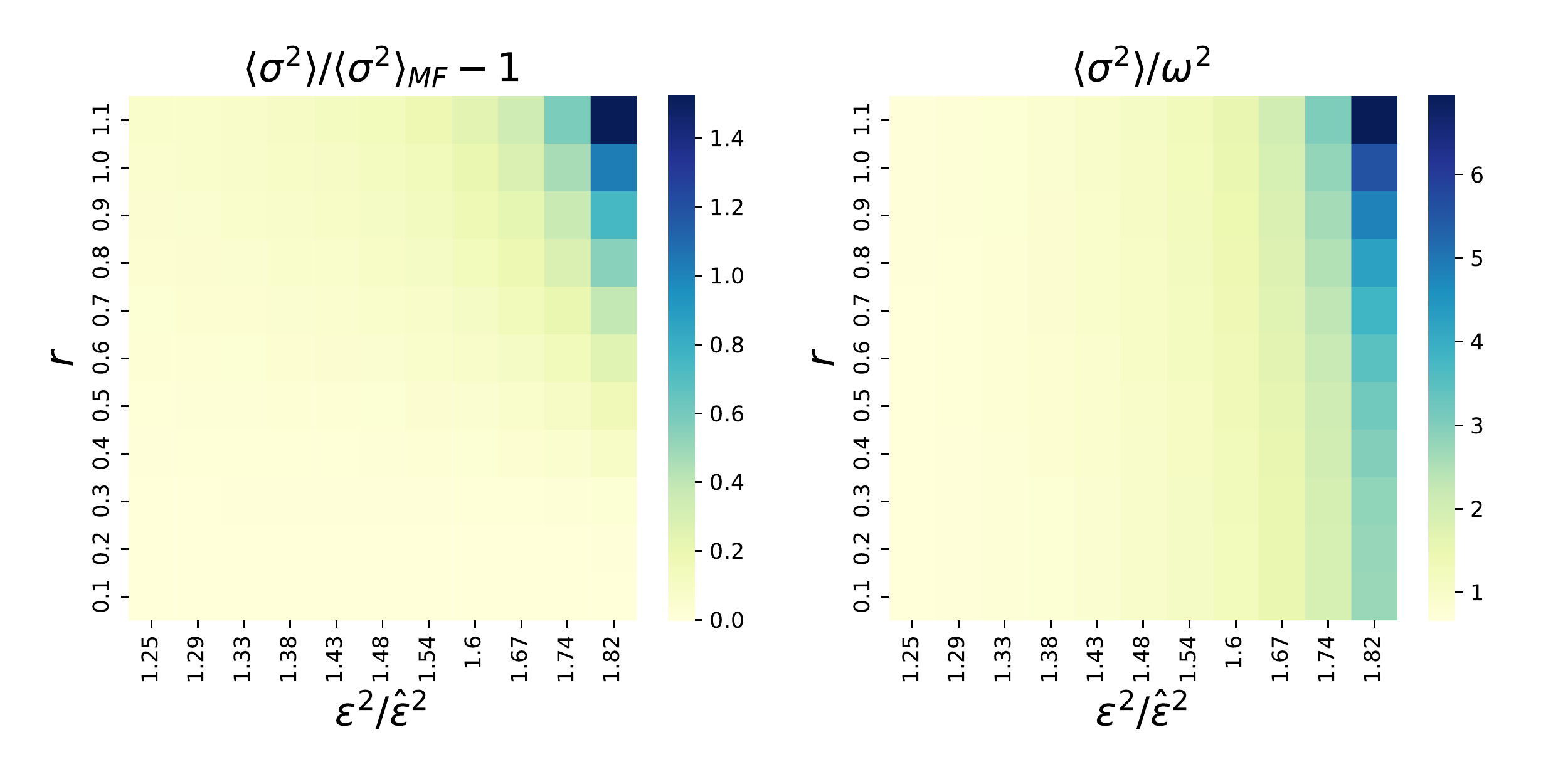}
    \caption{Mean excess variance as a function of the timescale ratio $r$ and of market maker's underestimation of noise trade variance, measured by $\epsilon^2/\hat\epsilon^2$. We set the variance of noise trade fluctuations to  $\delta\epsilon^2 /\epsilon^2= 0.1$. (Left) Normalized by the mean-field level. One can observe the departure from the mean-field description, due to the combined effect of fluctuations  and of market maker's underestimation of the noise trade variance. (Right) Mean excess variance. One can observe that a given level of excess variance can be obtained with different combinations of the parameters. }
    \label{fig:mean_excess}
\end{figure}

Consider a fluctuating noise trade variance modeled as an AR(1) process with positive and finite correlation timescale $\tau_{\NT}$, as the noise trade variance we defined in Eq.~\eqref{eq:NT_volatility}. 
 In principle, we can compute the mean value of excess variance in the sticky expectation regime of our model starting from Eq.~\eqref{eq:empirical_vol}. However, the correlation of the noise term complicates the analysis. Therefore, we consider the regime of fast vanishing ACF of noise trade variance fluctuations ($r\ll 1$) (or small overall level of fluctuations, i.e., $\delta\epsilon^2\ll 1$). In fact, since $\epsilon^2_k$ is Gaussian, all high order auto-correlations boil down to terms proportional to products of two time correlation functions between fluctuation $\delta\epsilon^2_k$ terms, which are proportional to $\delta\epsilon^2\exp(-1/r)\ll 1$. The summation to first order in $\delta\epsilon^2 \exp(-1/r)$ leads to a modified mean excess variance.
In particular, the mean excess variance is still of the form given by Eq.~\eqref{eq:MF_excess}, but the critical value of market maker's belief about noise trade variance is modified to:
\begin{equation}
    \label{eq:critical_value}
    \hat\epsilon^2_c = \left. \hat\epsilon^2_c\right|_{\MF} +\frac{\delta\epsilon^2}{2}\exp(-1/\tau_\NT).
\end{equation}
Similarly, the slow relaxation timescale $k_\text{slow}$ is given by Eq.~\eqref{MF_timescale}, where one has to replace the MF critical parameter with its `fluctuations aware' version given above.
These findings imply that a given level of mean excess variance can be obtained with a smaller underestimation of the market maker about noise trade variance if fluctuations are present, as we show in Figure \ref{fig:mean_excess}.
There, in the left panel, one can see that mean excess variance is equal to its MF value if $r$ is small (bottom part of the plot) or if  $\hat\epsilon^2$ is negligible (left part of the plot). Conversely, as $\hat\epsilon^2$ gets closer to the critical value (right part of the plot), one can see a sharp multiplicative increase as the fluctuations are increased (top-right corner in the plot). The increase of the mean excess variance as fluctuations are more persistent (larger $r$) can be interpreted by saying that  $1-\hat\epsilon^2_c/\hat\epsilon^2$ is getting smaller, implying an increase of the overall mean excess variance (see Eq.~\eqref{eq:MF_excess} properly modified by Eq.~\eqref{eq:critical_value}). The right panel of Fig.~\ref{fig:mean_excess} shows that a given level of price variance can be obtained with different choices of  market maker's underestimation about noise trade variance level and the persistence of noise trade variance fluctuations. 

While in Fig.~\ref{fig:mean_excess} a high mean excess variance requires in any case an underestimation of noise trade variance level, note that high excess variance can be obtained also if the market maker is right about the mean level value of fundamental price variance, but noise trade variance fluctuations are large in magnitude and are persistent. In this case, in fact, the situation our model describes is qualitatively  similar to that encountered in economic models where quasi-non-ergodicity is taken into account~\cite{quasi-non-ergodicity}; quasi-non-ergodicity occurs when a stochastic process is
ergodic at very long-time horizons, but where ergodicity breaks down on a time
scale at which realisations from the process might realistically be observed by a
human agent.
This is exactly the situation the market maker faces if $r<\infty$; in this case, in fact, his belief about the fundamental price variance is sensitive to noise trade variance fluctuations.

\paragraph{Numerical simulations}

\begin{figure}
    \includegraphics[scale = 0.63]{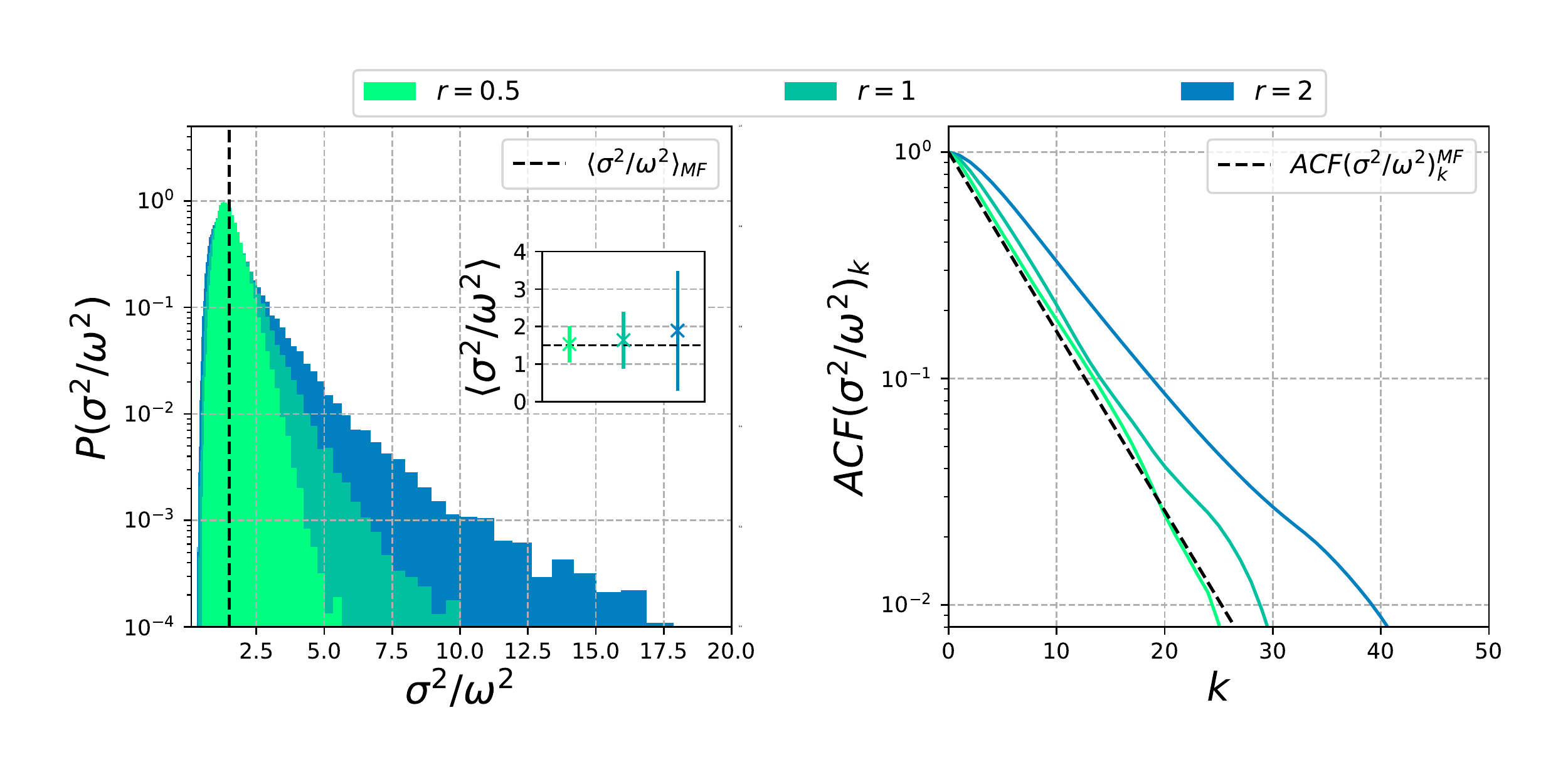}
    \caption{Simulations varying the timescale ratio $r$, while keeping fixed the updating time to  $\tau_{\text{rev}}= 1000$. We set $\delta\epsilon^2/\epsilon = 0.2$ and $\hat\epsilon^2/\epsilon^2 = 0.6$. (Left)~PDF of excess variance. Inset: MF level of excess variance compared with simulations outcomes (with error bars). (Right) ACF of excess variance. Black dashed lines are related to MF results in the sticky expectation regime of our model.\protect\footnotemark} 
    \label{fig:3}
\end{figure}

We analyse simulations with fixed  updating timescale $\tau_{\text{rev}}$ and different timescales of fluctuations of noisy order flow volatility $\tau_{\NT}$.
In the left panel of Fig.~\ref{fig:3}, one sees that the  more persistent the fluctuations (large $r$), the more the PDF  of excess  variance is skewed towards large values (the shift is related to the underestimation of the noise trade's level, as before).  The reason why this occurs can be understood from  Eq.~\eqref{eq:empirical_vol}: the more the noisy order flow volatility is serially correlated, the more the feedback dynamics on price volatility persists in the same direction, leading to large fluctuations of excess volatility. In the following section, we characterize analytically the tail behavior in the sticky expectation regime. The inset  shows that the more persistent the fluctuations of noisy order flow volatility, the higher the mean excess volatility, consistent with the results presented in  Fig.~\ref{fig:mean_excess}. 

\footnotetext{Note that we anticipated the MF result regarding the ACF of excess variance in Sec.~\ref{subsec:garc_comparison}.}

\subsection{Intermittent volatility dynamics}

In the following we characterise the intermittent  volatility dynamics. As in the previous section, first the sticky expectation regime is considered, and then our findings are further substantiated with the outcomes of the simulation of the model.

The intermittent dynamics of excess volatility can be characterized, at a first approximation, by the tail exponent of the Cumulative Distribution Function (CDF) of price volatility, and by the temporal decay of the ACF of the price variance, respectively  given by $\mu$ and $\tau_\text{ACF}$ (actually, to provide an accurate description of empirical findings, more than one timescale is needed to characterize the ACF of the price variance~\cite{bouchaud_risk}). 
Results regarding the CDF's power law tail and the structure of the ACF of price variance are available for Kesten processes with iid multiplicative noise in the stable regime, where $\langle \epsilon^2_k/(2\hat\epsilon^2 )\rangle <1$. Below, we recall these important results, and we highlight how they change when an AR(1) process is considered.
 
 If the  noise trade variance $\epsilon^2_k$  is iid, the CDF of excess volatility,  decays asymptotically as a power law with exponent $\mu$, which has the following form~\cite{multiplicative,condensation}:
 \begin{equation}
\label{eq:exponent}
\left\langle \left(\cfrac{\epsilon^2_k}{2\hat\epsilon^2}\right)^{\mu/2} \right\rangle = 1.
\end{equation}
Accordingly, the tail of the price volatility CDF is thicker, i.e., $\mu$ is smaller, the more the market maker underestimates the mean level noisy order flow volatility. 
Equation \eqref{eq:exponent} implies that the power law tail of the probability distribution shape is robust with respect to the underlying distribution of the multiplicative term $\epsilon^2_k/(2\hat\epsilon^2)$. This insensitivity to micro-structural details justifies  Kesten processes as an effective description of the universal  intermittent dynamics exhibited by price volatility. If $\epsilon^2_k$ are realizations of an AR(1) process, from numerical simulations we observe that the excess volatility has again an exponent $\mu$ which gets smaller with increasing persistence in noisy order flow volatility fluctuations.

If $\epsilon^2_k$ are iid, the long-time ACF of price volatility is  a single decaying exponential function. 
If $\mu>2$, which is the case for real markets~\cite{bouchaud_risk}, the correlation timescale $\tau_{\text{ACF}}$ of price volatility writes~\cite{condensation}:
\begin{equation}
\label{eq:constraint_Kesten}
\tau_\text{ACF} = \frac{8}{\mu-1}\left(\cfrac{\hat\epsilon}{\delta\epsilon}\right)^4 .
\end{equation}
Note that an interesting relation for $\mu$ can be obtained in the case of uncorrelated noisy order flow volatility fluctuations by comparing the equation above for $\tau_\text{ACF}$ with the one given in  Sec.~\ref{subsec:garc_comparison}.
According to the equation for $\tau_\text{ACF}$ given there, $\tau_\text{ACF}$ increases when $\hat\epsilon$ approaches the critical value $\epsilon_c$. From the equation above instead we conclude that $\tau_{\text{ACF}}$ increases if the level of fluctuations of noisy order flow volatility $\delta\epsilon$ decreases, recovering the MF regime we analysed in the previous section in the limit case where $\delta\epsilon = 0$. 
In the case where $\epsilon^2_k$ is an AR(1) process, the more persistent the noise trader's volatility fluctuations,  the more correlated price volatility, and the larger $\tau_\text{ACF}$. 

Empirical analysis conducted on the ACF of price volatility shows that at least two timescales are needed in order to capture its temporal structure~\cite{bouchaud_risk}. In the case where $\delta\epsilon^2_k$ is an AR(1) process, the ACF of price variance obtained with finite $r$ is captured by two decaying exponential functions, as we shall see below from numerical simulations.

\paragraph{Numerical simulations}
\label{subsec:power_tail}

\begin{figure}
    \centering
    \includegraphics[scale = 0.63]{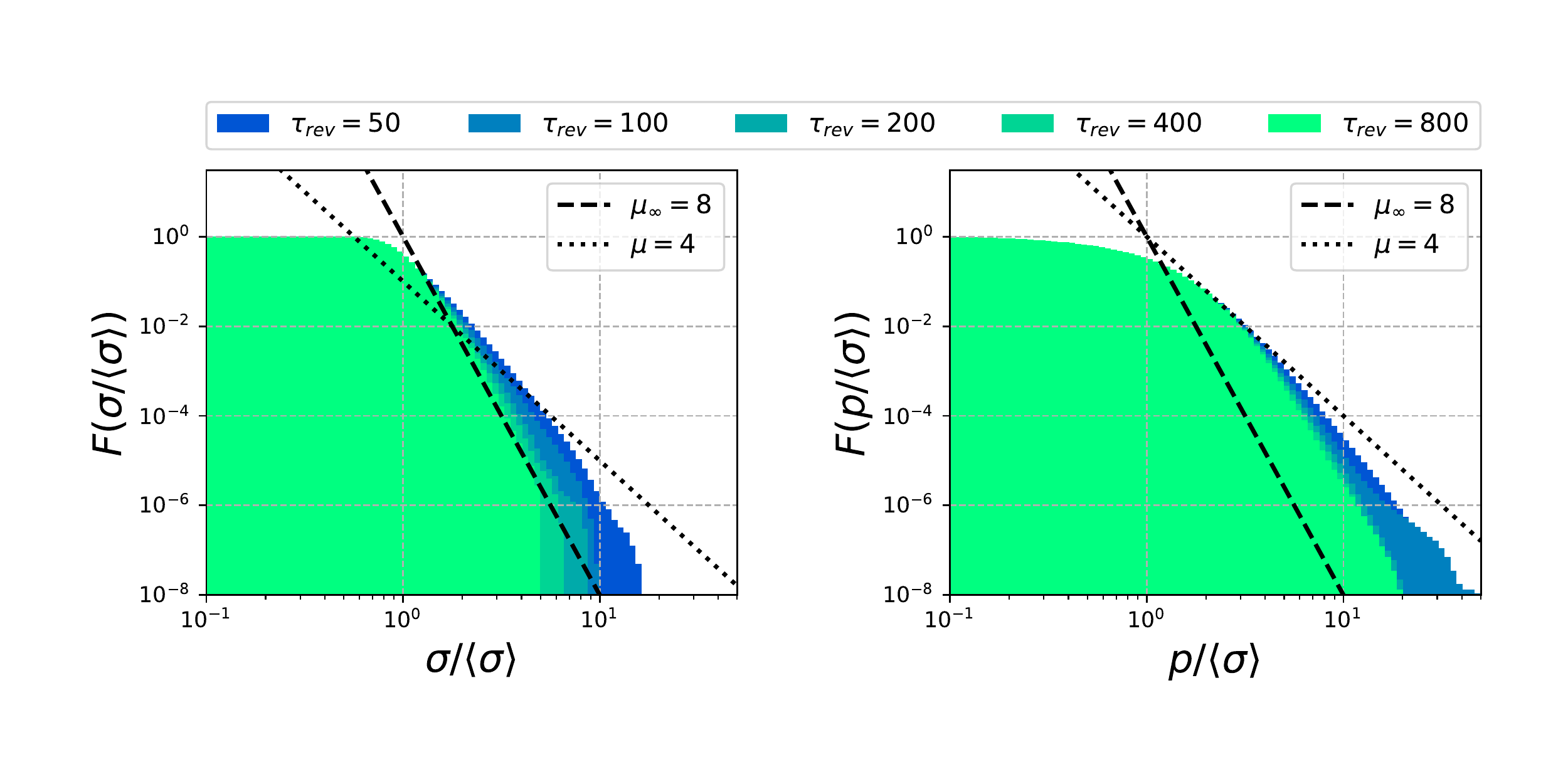}
    \caption{CDF of price volatility and price obtained from simulations with   $r = 1$, varying the updating timescale $\tau_{\text{rev}}$. The other parameters are chosen to be $\delta\epsilon^2 = 0.1$, $\hat\epsilon^2 = 0.52$. (Left) CDF of normalized price volatility. (Right) CDF of normalized price. Dotted line refer to results of the related Kesten process, while dashed line superposed for illustrative purposes. }
    \label{fig:comparison}
\end{figure}

Here we compare the results of the simulation of the model we presented in Sec.~\ref{sec:the_model} with the prediction of the Kesten process given by Eq.~\eqref{eq:exponent}. To do so, we run different simulations with fixed  ratio $r = 1$ approaching the limit $\tau_{\text{rev}}\rightarrow\infty$,  where the measurement error on the price volatility estimate  vanishes and the Kesten dynamics is recovered by construction. We set the model parameters in line with empirical predictions about mean excess volatility:  $\langle\sigma/\omega \rangle \sim 3.5$ is obtained if $ \hat\epsilon^2/\epsilon^2=0.52$ and $\delta\epsilon^2/\epsilon^2=0.1$. Interestingly, from Eq.~\eqref{eq:ratio_informed}, for every stock bought/sold by an informed trader, there are $\sim 25$  stocks bought/sold by the noise trader, which means that more than $90\%$ of the overall liquidity is made of non-informed trades.
 In the left panel of  Fig.~\ref{fig:comparison}, we show the CDF of normalized price volatility. A clear result is that the larger the updating timescale $\tau_{\text{rev}}$, the thinner the tail. This agrees with  intuition: the lower the measurement error on the price volatility estimate is, the lower the fluctuations of market maker's belief. The power law tail related to CDF of excess volatility obtained with $\tau_{\text{rev}} = 800$ is well approximated by $\mu_\infty =  8$, according to the result obtained from the simulation of the Kesten process, where the measurement error on the price volatility estimate is neglected since $\tau_\text{rev} \rightarrow \infty$. 
In the right panel of Fig.~\ref{fig:comparison}, we show that the power law tail of the price CDF does not change with respect to that related to price volatility, as expected. Let us mention here that one can obtain a thicker tail assuming a more realistic excess-demand  process, specifically a long range correlated process, typically with power law tail with exponent $1/2$~\cite{lillo}. 
Finally, the analysis of the ACF of price variance is done in the right panel of Fig.~\ref{fig:3}. One sees that the more persistent the noise trader fluctuations, that is, the larger $r$, the slower  the decay of the ACF  of price variance. This result can be explained again by recalling the analysis of the sticky expectation regime given in the previous section: the more correlated the multiplicative factor in Eq.~\eqref{eq:empirical_vol}, the more correlated price volatility as well. In the opposite limit, the results of the simulations match the mean-field result of the sticky expectation regime we analysed in the previous section, as expected. Note that the ACF for $r=2$ is clearly not captured by a single decaying exponential, as it is the case for empirical data~\cite{bouchaud_risk}.

\section{Extensions}
\label{sec:extensions}

\subsection{Risk-averse market maker}

It is well-known that risk-aversion of the market maker implies in Kyle-like frameworks a higher price volatility with respect to the risk-neutral case; see for example Ref.~\cite{Subrahmanyam}, where the case of an absolute relative risk-averse (CARA) market maker is considered. However, in order to explain the excess volatility encountered in empirical data an unrealistically high risk-aversion parameter has to be chosen~\cite{Leroy2013CanRA}. It is interesting to evaluate the importance of market maker's risk aversion in driving the mean price volatility in our framework. To do so, we modify the market maker's model behaviour, so that risk-aversion is enforced with CARA while the linearity of Eq.~\eqref{eq:linear_price_impact} is retained. The task of the market maker in this case is to choose $\Lambda_t$ such that:
\begin{equation}
\label{eq:cara}
    \mathbb{E}[U^\MM_t|q_t,\hat\omega_t,\hat\epsilon] = \mathbb{E}[q_t(p_t-p^\F_t)|q_t,\hat\omega_t,\hat\epsilon] -\rho_t \text{var}[q_t(p_t-p^\F_t)|q_t,\hat\omega_t,\hat\epsilon] = 0.
\end{equation}

The risk-averse market maker's strategy depends now also on a risk-aversion parameter $\rho_t$, in addition to the beliefs about noise trades and fundamental price volatility. Accordingly, the  self-consistent equation for the price impact function writes~\cite{Subrahmanyam}:
\begin{equation}
\label{eq:price_impact_dynamics_S}
    \Lambda_t =\frac{2\Lambda_{t-1}\hat\omega^2_t}{\hat\omega^2_t+4\Lambda^2_{t-1}\hat\epsilon^2}(1+2\Lambda_{t-1}\rho_t \hat\epsilon^2).
\end{equation}
It is standard in the quantitative finance literature to express the degree of risk-aversion in term of the Sharpe ratio, defined as the ratio between the expected gain and the square root of the risk associated to a given strategy. We define the Sharpe ratio per period as:
\begin{equation}
\label{eq:sharpe}
 S = \frac{\mathbb{E}[\mathbb{E}[q_t(p_t-p^\F_t)|q_t]]}{\sqrt{\mathbb{E}[\text{var}[q_t(p_t-p^\F_t)|q_t]]}}
\end{equation}
In order to have a constant Sharpe ratio per period, the market maker has to choose a risk-aversion coefficient  $\rho_t = \frac{S}{\hat\epsilon \hat\omega_t}$.

As in the case of a the risk-neutral market maker, the timescale needed for the price impact to reach the fixed point is still of order one, i.e.,  $\tau_{\eq} = 1$. The fixed point of Eq.~\eqref{eq:price_impact_dynamics_S} can be again computed, as we did in Sec.~\ref{subsec:fast}: assuming market maker's belief about the fundamental variance to be constant, $\hat\omega_t = \hat\omega$, the price impact and the expected price variance are respectively given by:
\begin{align}
\label{eq:pi_fixed_S}
    \Lambda_\infty &= \frac{S+\sqrt{1+S^2}}{2} \frac{\hat\omega}{\hat\epsilon},
    \\
\label{eq:vol_fixed_S}
    \hat\sigma^2_\infty &=    \frac{\hat\omega^2}{2}\left[1+S(S+\sqrt{1+S^2})\right].
\end{align}
The risk-aversion of the market maker increases the value of the price impact, and, consequently, it increases the value of the expected price volatility, leading, as we shall see, to an increase of the actual price volatility. 

The slow dynamics of excess variance in the sticky expectation regime defined by Eq.~\eqref{eq:scaling_limit} is again of the Kesten type. In fact, following the same steps which led to Eq.~\eqref{eq:empirical_vol}, one finds:
\begin{equation}
\frac{\sigma^2_{k}}{\omega^2} = \frac{1}{4}+\frac{\left(S+\sqrt{1+S^2}\right)^2}{2 (1+S(S+\sqrt{1+S^2}))}\frac{\epsilon^2_k}{\hat\epsilon^2}\frac{\sigma^2_{k-1}}{\omega^2}.
\end{equation}
The MF version of the equation above obtained with $\epsilon_t = \epsilon$, admits a positive finite fixed point only if $\hat\epsilon^2 > \epsilon^2_{c,S}|_\MF = \epsilon^2 \left(S+\sqrt{1+S^2}\right)^2/[2 (1+S(S+\sqrt{1+S^2}))]$. Note that $\epsilon^2_{c,S}\geq \epsilon^2_{c,S=0}$: therefore, if the market maker is risk-averse, the maximum error he can make (without preventing the stationary regime to establish) on the mean noisy order flow volatility is  lower than that of the risk-neutral case. The contribution to the overall price volatility due to the risk-aversion of the market maker is qualitatively similar to what we  investigated in Sec.~\ref{subsec:mean_excess}, where we compared the critical value of market maker's belief about noisy order flow volatility in presence or in absence of noisy order flow volatility fluctuations. 

\begin{figure}
    \centering
    \includegraphics[scale = 0.62]{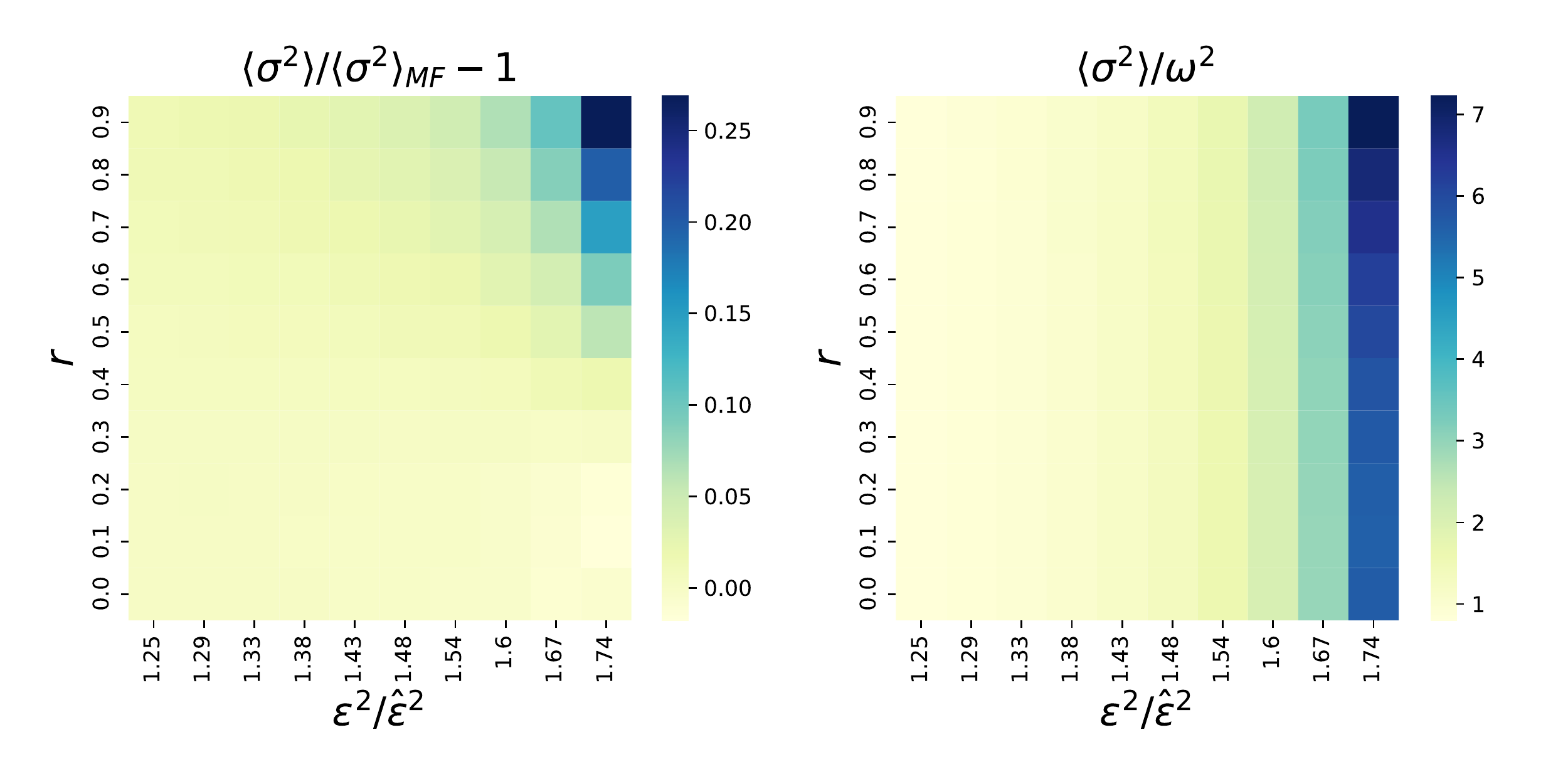}
    \caption{Mean excess variance as a function of the timescale ratio $r$ and of market maker's underestimation of noise trade variance, measured by $\epsilon^2/\hat\epsilon^2$ in the presence of a risk averse market maker with $S=0.1$. We set the variance of noise trade fluctuations to  $\delta\epsilon^2 /\epsilon^2= 0.15$. (Left) Normalized by the mean-field level. One can observe the departure from the mean-field description, due combined effect of fluctuations  and of market maker's underestimation of the noise trade variance. (Right) Mean excess variance. One can observe that a given level of excess variance can be obtained with different combinations of the parameters. The example analyzed in the main text correspond to the set of parameters identified by the bottom-right corners.}
    \label{fig:contourS}
\end{figure}

In the following we consider the  risk-averse case with realistic values of the parameters, simulating the dynamics in the sticky expectation regime equilibria with uncorrelated noisy order flow volatility. We do so by setting an equal value for the mean excess volatility compatible with empirical results reported in the literature~\cite{Shiller,Blackok}, i.e., $\langle \sigma/\omega \rangle \sim 2.5$.
For example, this level of excess volatility can be achieved in the risk-averse case with the choice of parameters $S = 0.1, \epsilon^2/\hat\epsilon^2 = 1.75$ and $ \delta\epsilon^2/\epsilon^2 = 0.15 $. We obtain a price volatility correlation timescale of $\tau_\text{ACF} \sim  25$ and an exponent $\mu \sim 5$ for the tail of the price volatilty CDF. Since in empirical works the ACF timescale of price volatility is of the order of months, when asking for a realistic value of the annualised Sharpe ratio (in a competitive market one would expect it to be of order 1), one obtains plausible values for the daily Sharpe ratio $S$ of the order of $\sim 0.1$.
This illustrative example highlights the fact that in order to describe a reasonably competitive market ($S\sim 0.1$), excess volatility cannot be accounted for solely on the basis of risk aversion, and needs to be justified by a large negative bias in the estimation of the average noise trade level $\hat\epsilon^2/\epsilon^2$, possibly boosted by the effect of the fluctuations. Indeed, one can explore the excess volatility for different values of $r$ as we did in Sec.~\ref{subsec:dynamic_market}. As it is shown in Fig.~\ref{fig:contourS}, the mean excess volatility increases if  the noise trade variance is correlated over time.

We thus provided a mechanism to generate realistic mean excess-volatility level while recovering an intermittent dynamics for price volatility which does not rely primarily on the risk-aversion of the market maker, providing a possible solution, in line with the AMH, for a long lasting problem in economic literature~\cite{Leroy2013CanRA}.

\subsection{Cost-averse noise trading, liquidity crises and flash crashes}
\label{subsec:cost}

Up to now we considered a passive noise trader, namely a noise trader who does not modify his trading intensity even if the price impact increases, resulting in higher trading costs. Moreover, the passive noise trader hypothesis implies that the level of liquidity is always finite and bounded from below by the constant liquidity provided by the noise trader. In reality, instead, the overall level of liquidity fluctuates over time;  the market can experience liquidity crises, that is,  situations in which the overall liquidity vanishes, while the price impact goes virtually to infinity. In what follows we consider a  cost-averse noise trader and we show how this modification can account for the  fragility exhibited by financial markets. 

 The noise trader demand $q^\NT_t$ minimises the expectation of a cost function $C^\NT_t$ made  of two terms: the first is the usual profit and loss term multiplied by the cost-aversion parameter $\phi>0$, while the second  the squared difference between the actual demand  $q^\NT_t$ and a (moving) trading target $q^\tgt_t$:
 \begin{equation}
    C^\NT_t = (q^\NT_t-q^\tgt_t)^2-\phi q^\NT_t p_t,
\end{equation}
where $q^\tgt_t$ are realisations of a Gaussian process with zero mean and time varying volatility $\epsilon_t$ which we assume to be known by the cost-averse noise trader.  The time varying volatility $\epsilon_t$ has an AR(1) structure. 
 We assume that, as was the case for the informed trader, the noise trader can infer past price impact functions and use the last known value in order to construct his strategy. Accordingly, the trading strategy of the cost-averse noise trader reads:
\begin{equation}
\label{eq:cost-averse_NT}
q^\NT_t = \frac{q^\tgt_t}{1+\phi\Lambda_{t-1}},
\end{equation}
We can relate the cost-aversion parameter $\phi$ with a tracking error $\xi>0$ which measures how much the noise trader can afford to be off with respect to his trading target. The squared tracking error $\xi^2$ is defined as follows:
\begin{equation}
\label{eq:epsilon}
    \xi^2 = \frac{\langle (q^\NT_t-q^\tgt_t)^2 \rangle_{\infty}}{\epsilon^2}
\end{equation}
where the subscript $\infty$ means that the noise trader sets his tracking error $\xi$ assuming that the price impact is equal to its fixed point value, which we assume can be computed by the noise trader. The link between the cost-aversion parameter $\phi$ and the tracking error parameter $\xi$ is given by $\phi = 2 \xi \hat\epsilon/\hat\omega$.

We suppose that the risk-neutral market maker does not know the volatility of the noise trader's target, but he has a prior about it, namely,  $\hat\epsilon$. Moreover, we assume that the market maker knows the cost-aversion parameter $\phi$ (or, equivalently, $\xi$) of the noise trader, for simplicity. The dynamics of the price impact function is therefore again given by Eq.~\eqref{eq:dynamical_syst_pre} where one has to take into account the new expression for the noise trade variance belief which stems from Eq.~\eqref{eq:cost-averse_NT}; this amounts to employ the substitution  $\hat\epsilon^2 \rightarrow \hat\epsilon^2/(1+\phi \Lambda_{t-1})^2$ in Eq.~\eqref{eq:dynamical_syst_pre}.
Accordingly, the long-time price impact function reads
\begin{align}
\label{eq:pi_fixed_cost}
    \Lambda_\infty &= \frac{1}{2(1-\xi)}\frac{\hat\omega}{\hat\epsilon},
\end{align}
while the expected price volatility is given again by Eq.~\eqref{eq:vol_fixed}. Note that the price volatility is not affected by the noise trader's cost aversion $\phi$ (or $\xi$) in the case where the market maker knows with absolute precision this parameter.
We assume that these fixed points are known by the noise trader, who uses such information to construct his strategy based on Eq.~\eqref{eq:epsilon}. The noise trade  variance at the fixed point is  $(1-\xi)^2\epsilon^2_t$ and vanishes if the noise trader's cost aversion is high, i.e., in the limit $\xi \rightarrow 1$;  accordingly, also the overall liquidity scales with $(1-\xi)^2$. Therefore, the higher the cost-aversion of the noise trader, the smaller is the overall level of liquidity in the market. At the same time, the price impact function diverges as $\xi \rightarrow 1$ such that the price volatility remains constant.

Interestingly, the cost-aversion of noise trades results in a kind of friction force which delays the approach of the price impact function to the fixed point. In fact, we find that the relaxation time related to the price impact dynamics with constant belief is given by:
\begin{equation}
\label{eq:eq_timescale_dep}
    \tau_\xi  = \frac{1}{1-\xi}.
\end{equation}
An important difference with the case of the passive noise trader, is that the relaxation timescale $\tau_\xi$ is now $\xi$-dependent. In particular, it diverges as the noise trader becomes extremely cost-averse ($\xi \rightarrow 1$), resulting in a market which ever evolves in a strongly out of equilibrium regime, where the liquidity vanishes, the price impact diverges,  while price volatility remains bounded. 

\begin{figure}
    \centering
    \includegraphics[scale = 0.55]{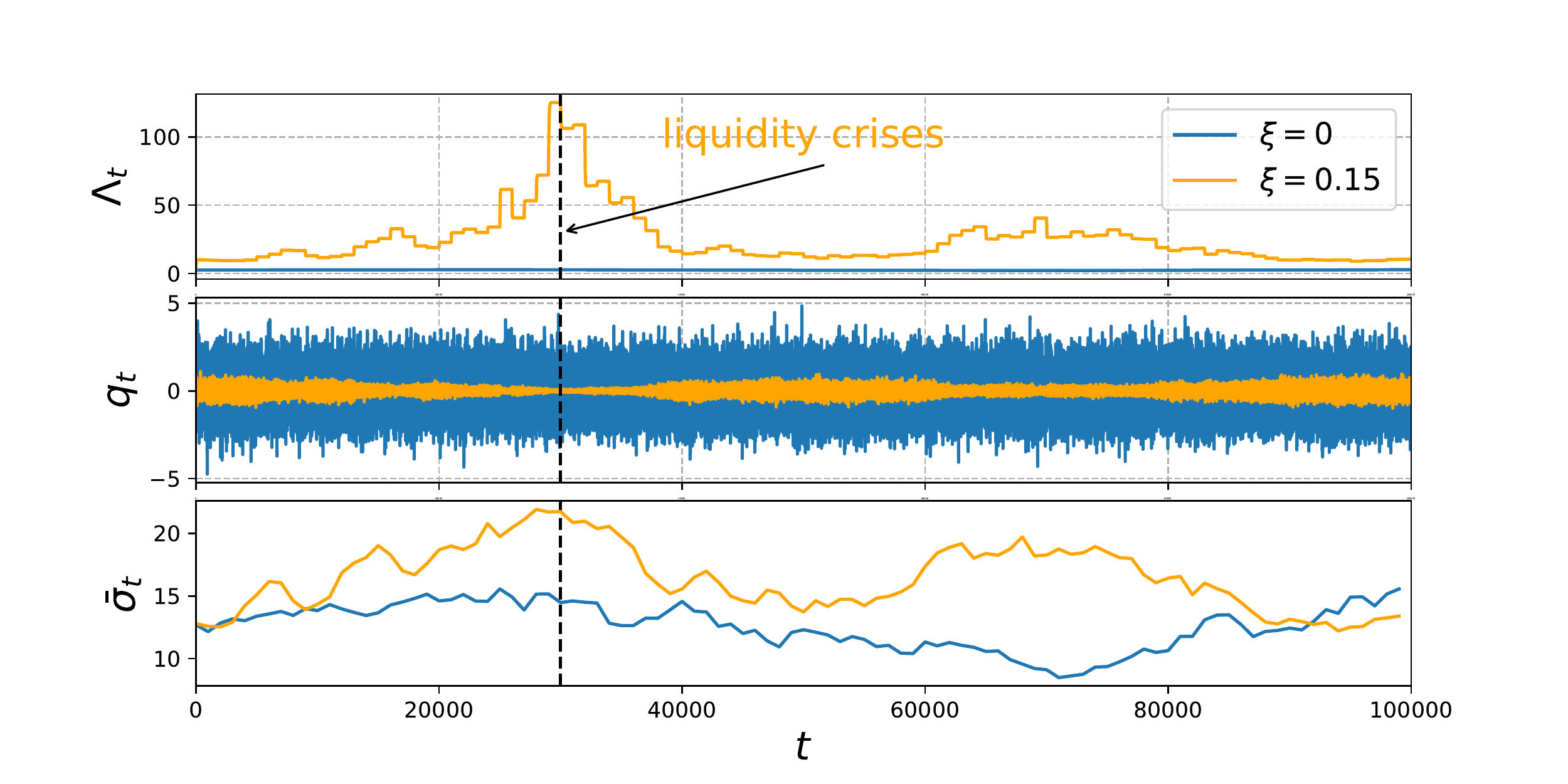}
    \caption{Comparison between sticky expectation regime dynamics with cost-averse and cost-neutral noise trader, with parameters $\hat\epsilon^2/\epsilon^2 = 0.52,\delta\epsilon^2/\epsilon^2 = 0.15$ and $\tau_\NT = 0$. (Top) price impact dynamics. (Middle) Total excess-demand dynamics. (Bottom) Price volatility estimate. Blue lines are related to cost-neutral noise trading, while orange lines are related to the cost-averse case. The black dashed line points at the regime where, in the case of cost-averse noise trader, the liquidity vanishes.}
    \label{fig:liquidity}
\end{figure}

The overall liquidity fluctuates much more than what was implied by the first version of our model, where the strategy of the noise trader was cost independent. We present this finding in Fig.~\ref{fig:liquidity}, where we show the results of the modified model simulation in the sticky expectation regime given by Eq.~\eqref{eq:scaling_limit}. One can see in blue the results of simulations where the noise trader is cost neutral, while in orange one sees the results where the noise trader is cost averse. It is clear that orange lines represent a regime where the liquidity and the price impact dynamics are more intertwined. In particular, the peak in the price impact function (orange line in the top panel) corresponds to the period in which the overall liquidity is small. Regarding the price volatility in presence of the cost-averse noise trader (orange line in bottom panel), while it  rises in the proximity of the  liquidity crises, the values attained are still comparable with those in absence of the noise trader's cost aversion. In fact, as we highlighted above, the equation for the price volatility does not change with respect to the case of cost-neutral noise trader if the market maker knows exactly the cost aversion parameter of the noise trader. 

To obtain a volatility profile that reacts to liquidity crises one has to relax the assumption that the market maker knows the true cost-aversion parameter of the noise trader $\xi$. With this modification, in fact, the resulting   excess-volatility dynamics in the sticky expectation regime reads
\begin{equation}
    \label{eq:price_vol_cost}
    \frac{\sigma^2_{k}}{\omega^2} = \frac{1}{4}\left[1+ \left(\frac{1-\xi^2_k}{1-\hat\xi^2}\right)^2 \frac{\epsilon^2_k}{2\hat\epsilon^2} \frac{\sigma^2_{k-1}}{\omega^2} \right],
\end{equation}
where $\hat\xi$ denotes the market maker belief about the true tracking error parameter of the noise trader.
The equation above shows that if the market maker overestimates the noise trader's cost aversion, and if $\hat\xi \rightarrow 1$, then the fixed point price volatility diverges, simulating a flash crash where the liquidity vanishes while the price volatility diverges.

\section{Conclusion and outlook}
\label{sec:conclusion}

Let us summarise what we have achieved. 
The aim of this paper was to modify the classic framework of asymmetrically informed agents in presence of noise in order to capture the  excess volatility and  volatility clustering exhibited by real financial markets, without resorting to unrealistic risk-averse agents~\cite{Leroy2013CanRA} nor fundamental innovation clustering. 
We proposed a modification of the paradigmatic Kyle model of price formation, where agents adapt to the ever-evolving market conditions. Accordingly, each trader has its own model of reality, which does not generally match that of other traders; for example, we assumed that the market maker sets prices, while he does not know the precise level of signals and noise, that is, of the fundamental price and the non-informed trades variance. Moreover, we allowed the market maker to update his belief about the unobservable fundamental price variance by comparing the price variance expectations with empirical estimates. The market maker therefore acts based on a system of temporarily fulfilled expectations~\cite{Arthur}. 
We analytically characterised the model in a realistic limit: the resulting stationary dynamics of excess volatility is of the Kesten type~\cite{kesten}, i.e. a stochastic multiplicative process repelled from zero~\cite{multiplicative}, which exhibits intermittent dynamics and power law tails.
Interestingly, power law behaviour is robust against changes in the way market conditions evolve across time.  This important finding is in line with the idea that adaptive behaviour in the presence of noise, being a universal feature of human behaviour, can be reflected in the universality of price dynamics across time and across markets.  
As a side, yet compelling, result, the microfoundation we propose is able to rationalise GARCH models while taking into account the relation of the price with fundamentals; as a consequence, the parameters that define the GARCH process are directly linked with non-observable quantities such as volatility of fundamentals and of noise trades. Hence, the microfoundation we propose enhances the interpretability as well as the predictive capacity of GARCH models. Our model predicts that some excess volatility can be accounted for by a mechanism based on quasi-non-ergodicity which has been recently proposed as a way to overcome the classic strong rationality paradigm~\cite{quasi-non-ergodicity}; in fact, we have shown that excess volatility is higher in situations where the updating timescale of the market maker is of the same order of magnitude as the timescale with which the market conditions vary.  

The microfoundation we propose seems suitable to describe the large fraction of price jumps not directly linked to fundamental innovations' arrivals. In fact, it predicts symmetric price jumps, in line with  empirical findings~\cite{Marcaccioli}; therefore, it points at the fact that symmetric GARCH models are more prone to describe price jumps not related to external fundamental innovations, at odds with the EMH story which relies on fundamental innovation clustering.

The present framework is versatile. For instance, in Sec.~\ref{subsec:cost} we chose to analyse the case of a cost-averse noise trader, and showed that, the more  cost-aversion there is, the more  fragile and illiquid the market is. If the market maker does not know precisely the noise trader's cost-aversion, flash crashes can occur. Yet, we analysed only the case of constant cost-aversion. It could be of real interest to couple the cost-aversion of the noise trader and the updating timescale of the market maker's model with the current price volatility level. This coupling should lead to a more realistic description of the fragile and highly intertwined endogenous dynamics able to account for flash crashes in financial markets~\cite{fosset}. 
Another relevant extension, that may interest researchers dealing with financial contagion~\cite{Gualdi_systemic_risk,poledna_financial_contagion}, is to consider a multi-asset generalisation. In fact, by assuming a network of interdependent fundamental prices, one can explore the extra fragility due to traders' interactions.
Finally, one could think of an extension with a correlated excess-demand process, in line with empirical observations~\cite{lillo}. See Refs.~\cite{Vodret_2021, Vodret2022,cordoni} for inspiration on how to tackle  such ideas.

\section*{Acknowledgments}
We warmly thank F. Moret who contributed to  the early stages of the analysis with a cost-averse noise trader, as well as J.-P. Bouchaud, C. H. Hommes, R. Marcaccioli and Y. Zhang  for interesting discussions. This research was conducted within the Econophysics \& Complex Systems Research Chair, under the aegis of
the Fondation du Risque, the Fondation de l’Ecole polytechnique, the Ecole polytechnique and Capital Fund
Management.

\bibliographystyle{apsrev4-1}
\bibliography{references_updated}

\clearpage

\appendix

\section{How to simulate the model}
\label{app:how_to_simulate}
Below we present a pseudo-code to perform simulations of the model presented in Sec.~\ref{sec:the_model}.

\begin{algorithm}

Consider a simulation which starts from $t=1$ and $k=0$, with  initial conditions  
$\{\Lambda_0, \hat\omega_0, \delta\epsilon_1\}$ and 

total duration  $T = \tau_{\text{rev}} K$, with $K$ a positive integer. The other parameters needed to run the

simulations are  $\{\omega,\epsilon,\delta\epsilon,\hat\epsilon,\tau_{\NT},\tau_{\text{rev}}\}$. We set $\omega = \epsilon = 1$.

\vspace{0.5cm}

\For{$t\leq T$}{


    Build the excess demand $q_t = q^\IT_t + q^\NT_t $: 
    
    the informed trade $q^\IT_t$ is given by  Eq.~\eqref{eq:IT_strategy}, i.e.,  it is the ratio between a realization of the 
    
    fundamental price $p^\F_t$, obtained from a Gaussian process with zero mean and volatility $\omega$, and 
    
    the last observed price impact function $\Lambda_{t-1}$; 
    the noise trade $q^\NT_t$ is a realization of a Gaussian 
    
    process with zero mean and volatility  $\epsilon_t$. 
    Then, the price $p_t$ is given by   Eq.~\eqref{eq:linear_price_impact} and Eq.~\eqref{eq:dynamical_syst_pre}, 
    with 
    
    fundamental price and noisy order flow volatility beliefs respectively given by $\hat\omega_{k \tau_{\text{rev}}}$ 
    and $\hat\epsilon$; 
    
    in doing this last step, one updates the price impact function.

\vspace{0.4cm}

    \If {$t \neq  (k+1) \tau_\text{rev}$}{
       $t++$. 
        
        Update noisy order flow volatility $\epsilon_t$ according to Eqs.~\eqref{eq:NT_volatility} and \eqref{eq:errrorAR1}.}
     
     \Else{
     First, construct a price volatility estimate $\bar\sigma_{k \tau_\text{rev}}$ from the past $\tau_\text{rev}$ prices $p_t$.
     
     Then, update the fundamental price volatility $\omega_{k\tau_\text{rev}}$ according to Eq.~\eqref{eq:update_fundamental}. 
$t++,k++$.}}\end{algorithm}

Note that the simulation is  characterized by four ($+1$ timescale, if the market is stable) timescales: 
\begin{itemize}
    \item[1.] $t$, over which trading take place.
     \item[2.] $\tau_{\NT}$, over which noisy order flow volatility fluctuates. This is a parameter which the modeler has to fix.
    \item[3.]  $\tau_\text{fast}$, over which the fast dynamics of price impact reaches the stationary regime. This parameter controls the fast dynamics of price impact, given by Eq.~\eqref{eq:dynamical_syst_pre}, which has been analyzed in Sec.~\ref{subsec:fast}.
    \item[4.] $\tau_{\text{rev}}$, over which the market maker updates his  belief about  fundamental price volatility. This is a parameter which the modeler has to fix.
    \item[5.] $\tau_{\text{slow}}$, over which the belief of the market maker converge in distribution, if the market is stable. In this case, this parameter controls the slow dynamics of market maker's belief, given by Eq.~\eqref{eq:update_fundamental} analyzed in Sec.~\ref{subsec:slow}.

\end{itemize}   

\end{document}